\documentclass{article}

\usepackage[nonatbib, final]{svrhm_2021}

\usepackage[numbers]{natbib}
\usepackage[utf8]{inputenc} % allow utf-8 input
\usepackage[T1]{fontenc}    % use 8-bit T1 fonts
\usepackage{hyperref}       % hyperlinks
\usepackage{url}            % simple URL typesetting
\usepackage{booktabs}       % professional-quality tables
\usepackage{amsfonts}       % blackboard math symbols
\usepackage{nicefrac}       % compact symbols for 1/2, etc.
\usepackage{microtype}      % microtypography
\usepackage{xcolor}         % colors
\usepackage{amsmath}
\usepackage{amssymb}
\usepackage[utf8]{inputenc}
\usepackage{dsfont}
\usepackage{lipsum}
\usepackage{tikz}
\usetikzlibrary{positioning}
\usepackage{graphicx}
\usepackage{wrapfig}
\usepackage{cleveref}
\usepackage{float}
\PassOptionsToPackage{numbers}{natbib}

\newcommand{\ncite}[1]{[\citenum{#1}]}

\title{Modeling Category-Selective Cortical Regions with Topographic Variational Autoencoders}

\author{%
  T. Anderson Keller\thanks{Equal contribution. Work done while Qinghe Gao was interning at UvA-Bosch Delta Lab.} \\
  UvA-Bosch Delta Lab\\
  University of Amsterdam\\
  \And
  Qinghe Gao\footnotemark[1] \\
  ChemEngAI Lab\\
  Delft University of Technology\\
  \And
  Max Welling\\
  UvA-Bosch Delta Lab\\
  University of Amsterdam\\
}

\begin{document}

\maketitle

\begin{abstract}
Category-selectivity in the brain describes the observation that certain spatially localized areas of the cerebral cortex tend to respond robustly and selectively to stimuli from specific limited categories. One of the most well known examples of category-selectivity is the Fusiform Face Area (FFA), an area of the inferior temporal cortex in primates which responds preferentially to images of faces when compared with objects or other generic stimuli. In this work, we leverage the newly introduced Topographic Variational Autoencoder to model the emergence of such localized category-selectivity in an unsupervised manner. Experimentally, we demonstrate our model yields spatially dense neural clusters selective to faces, bodies, and places through visualized maps of Cohen's d metric. We compare our model with related supervised approaches, namely the Topographic Deep Artificial Neural Network (TDANN) of \citet{TDANN}, and discuss both theoretical and empirical similarities. Finally, we show preliminary results suggesting that our model yields a nested spatial hierarchy of increasingly abstract categories, analogous to observations from the human ventral temporal cortex. 
\end{abstract}

\section{Introduction}
Category-selectivity is observed throughout the cerebral cortex. At a high level it describes the observation that certain localized regions of the cortical surface have been measured to respond preferentially to specific stimuli when compared with a set of alternative control images. It has been measured across a diversity of species \cite{ffa, monkeyffa, monkeyppa}, directly through fMRI and neural recordings \cite{monkeyfmri}, and more indirectly through observational studies of patients with localized cortical damage \cite{lesionPPA}. Examples of category-selective areas in the visual stream include the Fuisform Face Area (FFA) \cite{ffa}, the Parahippocampal Place Area (PPA) \cite{ppa, monkeyppa}, and the Extrastriate Body Area (EBA) \cite{eba} which respond selectively to faces, places, and bodies respectively. However, the extent of category-selectivity does not stop at such basic categories. Instead, selective maps have been observed for both more abstract `superordinate' categories, such as animacy versus inanimacy \cite{animacy, size}, as well as for more fine-grained `subordinate' categories such as human-faces versus animal-faces \cite{hierarchy}. These maps are seen to be superimposed on one-another such that the same cortical region expresses selectivity simultaneously to animate objects and human-faces, while other spatially disjoint regions are simultaneously selective to inanimacy and `places' (images of scenes). Such overlapping maps have been interpreted by some researchers as nested hierarchies of increasingly abstract categories, potentially serving to increase the speed and efficiency of classification \cite{stanfordsurvey}. 

In interpreting these observations, one may naturally wonder as to the origins of such localized  specialization. From the current literature, the driving factors can roughly be divided into two potentially complimentary categories: anatomical, and information theoretic. Anatomically, the arrangement and properties of different cell bodies can be observed to vary slightly in different regions of the cortex in loose alignment with category selectivity \cite{mfs, receptors, anatomy}, possibly serving as an innate blueprint for specialization. In the same category, the principle of `wiring length minimization' \cite{koulakov2001orientation, tension} posits that evolutionary pressure has encouraged the brain to reduce the cumulative length of neural connections in order to reduce the costs associated with the volume, building, maintenance, and use of such connections. Computational models which attempt to integrate such wiring length constraints \cite{TDANN, VTCSOM, Blauch2021} have recently have been observed to yield localized category selectivity such as `face patches' similar to those of macaque monkeys. 
A hypothesized second factor behind the emergence of category specialization, which has recently gained increasing empirical support, derives its explanatory power from information theory. Empirical studies have discovered that sufficiently deep convolutional neural networks naturally learn distinct and largely separate sets of features for certain domains such as faces and objects. Specifically, the work of \citet{Dobs2021}, showed that feature maps in the later layers of deep convolutional neural networks can be effectively segregated into object and face features such that lesioning one set of feature maps does not significantly impact performance of the network on classification of the other data domain. Such experiments, and others \cite{functional_spec, konkle_selfsup}, suggest that the specialization of neurons may simply be an optimal code for representing the natural statistics of the underlying data when given a sufficiently powerful feature extractor. % This idea of an 'optimal code' is reminiscent of the principle of redundancy reduction, or the efficient coding hypothesis \cite{barlow}, upon which the model we present in this paper is primarily founded.  

Pursuant to these ideas, this work proposes that a single underlying information theoretic principle, namely the principle of redundancy reduction \cite{barlow1961possible}, may account for localized category selectivity while simultaneously serving as a principled unsupervised learning algorithm. Simply, the principle of redundancy reduction states that an optimal coding scheme is one which minimizes the transmission of redundant information. Applied to neural systems, this describes the ideal network as one which has statistically maximally independant activations -- yielding a form of specialization. This idea served as  the impetus for computational frameworks such as Sparse Coding \cite{olshausen1997sparse} and Independant Component Analysis (ICA) \cite{SejnowskiICA, comon1994independent, hyvarinen1998independent, hyvarinen2000independent}. Interestingly, however, further work showed that features learned by linear ICA models were not entirely independant, but indeed contained correlation of higher order statistics. In response, researchers proposed a more efficient code could be achieved by modeling these residual dependencies with a hierarchical topographic extension to ICA \cite{hyvarinen2001topographic, HYVARINEN20012413}, separating out the higher order 'variance generating' variables, and combining them locally to form topographically organized latent variables. 
Such a framework shares a striking resemblance to models of divisive normalization \cite{simoncelli, GDN}, but inversely formulated as a generative model. 
Ultimately, the features learned by such models were reminiscent of pinwheel structures observed in V1, encouraging multiple comparisons with topographic organization in the biological visual system \cite{hyvarinen2009natural, HYVARINEN20012413, ma2008overcomplete}. 

In this work, we leverage the recently introduced Topographic Variational Autoencoder \cite{TVAE, PCTVAE}, a modern instantiation of such a topographic generative model, and demonstrate that it is capable of modeling localized category selectivity as well as higher order abstract organization, guided by a single unsupervised learning principle. We quantitatively validate category selectivity through visualization of Cohen's d effect size metric for different image classes, showing selective clusters for faces, bodies, and places. We compare our model with another recently developed model of topographic organization%based on a supervised wiring cost proxy
, the Topographic Deep Artificial Neural Network (TDANN) \cite{TDANN}, and demonstrate qualitatively similar results with an unsupervised learning rule. Finally, we show preliminary results indicating that our model contains a nested spatial hierarchy of increasingly abstract categories, similar to those observed in the human ventral temporal cortex  \cite{hierarchy, stanfordsurvey}.

% \input{sections/relatedwork}

% \section{Background}
% In this section we will briefly review the class of topographic generative models, and further demonstrate how such models can be efficiently trained through variational inference. 
\vspace{-1mm}
\section{Related Work} 
\vspace{-1mm}
Recently, a number of models of topographic organization in the visual system have been developed leveraging modern deep neural networks. \citet{VTCSOM} demonstrated category-selective regions, as well as a nested spatial hierarchy of selectivity, through the use of self-organizing maps (SOMs). Due to the challenges with scaling SOMs,  the inputs were dimensionality-reduced with PCA, limiting the applicability of the algorithm to arbitrary neural network architectures. Concurrently with our work, \citet{Blauch2021} developed the Interactive Topographic Network (ITN), inducing local correlation through locally-biased excitatory feedforward connections in a biologically-constrained model. Most related to our work, the TDANN of \citet{TDANN} incorporated a biologically derived proxy for wiring length cost into the fully connected layers of a supervised Alexnet model \cite{alexnet}, and similarly demonstrated emergent localized category-selectivity. Our model differs from these in that it  explicitly formulates a properly normalized density over the input data with topographic organization originating as a prior over latent the variables -- thereby unifying feature extraction and topographic organization into a single training objective: maximization of the data likelihood.

\section{Background}
\paragraph{Variational Autoencoders} 
Bayesian modeling, the theoretical framework underlying probabilistic generative models, has been proposed in multiple studies as a potential model of human learning \cite{TENENBAUM2006309, Tenenbaum1998BayesianMO}. Abstractly, the goal of a generative model is to accurately capture the true data generating process. Latent variable models propose to achieve this by defining a joint distribution over observations $\mathbf{X}$ and unobserved latent variables $\mathbf{T}$, commonly assuming that the joint factorizes into the product of a conditional `generative' distribution and a prior: $p_{\mathbf{X}, \mathbf{T}}(\mathbf{x}, \mathbf{t}) = p_{\mathbf{X}| \mathbf{T}}(\mathbf{x}|\mathbf{t})p_{\mathbf{T}}(\mathbf{t})$. These distributions are often parameterized with deep neural networks, earning the title `Deep Latent Variable Models'. The goal of training such  generative models is then to maximize the marginal likelihood of the data, $p_{\theta}(\mathbf{x}) = \int p_{\theta}(\mathbf{x}, \mathbf{t}) d\mathbf{t}$, with respect to the parameters $\theta$. However,  due to the intractability of the integral and the true posterior $p_{\theta}(\mathbf{t}|\mathbf{x})$, this is almost always intractable to compute exactly. Approximate solutions such as the Variational Autoencoder (VAE) \cite{kingma2013auto} were thus developed to provide tractable lower bounds on the data likelihood. Simply, in the VAE framework, an approximate posterior $q_{\phi}(\mathbf{t}|\mathbf{x})$ for the latent variables is separately parameterized and optimized to be close to the true posterior through the Evidence Lower Bound (ELBO):
\[\mathcal{L}_{\theta, \phi}(\mathbf{x}) = \mathbb{E}_{q_{\phi}(\mathbf{t}|\mathbf{x})} \left(\log p_{\theta}(\mathbf{x}|\mathbf{t}) - D_{KL}[q_{\phi}(\mathbf{t}|\mathbf{x}) || p_{\mathbf{T}}(\mathbf{t})]\right) \leq \log p_{\theta}(\mathbf{x})\] Through use of the reparameterization trick \cite{kingma2013auto, rezende2014stochastic}, all parameters $\phi$ and $\theta$ can then be simultaneously optimized with stochastic gradient descent. 

\paragraph{Topographic Generative Models} 
 In standard generative models such as ICA or Variational Autoencoders (VAEs) \cite{rezende2014stochastic, kingma2013auto}, it is common to define a prior over latent variables such that all variables are independant. Topographic generative models differ from this by instead having a more complex correlation structure defined by the spatial distance between variables in a pre-defined topographic layout. In Topographic ICA (TICA) \cite{hyvarinen2001topographic}, such a local-correlation structure was shown to be efficiently achievable through a 2-layer hierarchical generative model. Specifically, at the highest layer, a set of `variance generating' variables $\mathbf{V}$ are independently sampled and subsequently summed in local neighborhoods to determine the variance of lower level topographic variables $\mathbf{T}$. Formally, for independant normal variables $\mathbf{V} \sim \mathcal{N}(\mathbf{0}, \mathbf{I})$, the variances of $\mathbf{T}$ are given by a non-linearity $\phi$ applied to local sums of $\mathbf{V}$: $\boldsymbol{\sigma} = \phi(\mathbf{W}\mathbf{V})$, where we have expressed the local-sum operation in matrix form here for conciseness. A well known example of such a `local-sum' matrix $\mathbf{W}$ is the matrix representation of the convolution operation. The vector $\mathbf{T} \sim \mathcal{N}(\mathbf{0}, \boldsymbol{\sigma}^2 \mathbf{I})$ can then be seen to have correlations of variance across elements $T_i$ \& $T_j$ if these elements are `connected by $\mathbf{W}$', and thus share a subset of their variance generating variables $V$. 
Interestingly, the proposed wiring length proxy employed in the TDANN  \cite{TDANN} turns out to be based on the same underlying statistical property as the topographic generative models described in this work, namely local correlation. This suggests that these ideas may not be mutually exclusive, and hints at a potential fundamental connection between wiring length minimization and a generative modeling perspective of the brain.

% \subsection{The Topographic Variational Autoencoder}
\paragraph{The Topographic VAE} Inspired by linear topographic generative models such as TICA, the Topographic Variational Autoencoder (TVAE) \cite{TVAE} was recently introduced to train deep \emph{nonlinear} latent variable models with topographic structure. The model places a Topographic Product of Student's-T prior \cite{welling2003learning, Osindero2006} over the latent variables, and achieves efficient training through a hierarchical construction identical to that of TICA. Formally, the model parameterizes the conditional generative distribution with a powerful function approximator $p_{\theta}(\mathbf{x} | g_{\theta}(\mathbf{t}))$, and trains the paramters of this model through the use of \emph{two} approximate posteriors $q_{\phi}(\mathbf{z}|\mathbf{x})$ and $q_{\gamma}(\mathbf{u}|\mathbf{x})$ which are combined to construct the topographic $\mathbf{t}$ variable. Explicitly: 
% The Topographic VAE
\begin{gather}
\label{eqn:tvae1}
        q_{\phi}(\mathbf{z}|\mathbf{x}) = \mathcal{N}\big(\mathbf{z}; \mu_{\phi}(\mathbf{x}), \sigma_{\phi}(\mathbf{x}) \mathbf{I}\big) \hspace{8mm} q_{\gamma}(\mathbf{u}|\mathbf{x}) = \mathcal{N}\big(\mathbf{u} ; \mu_{\gamma}(\mathbf{x}), \sigma_{\gamma}(\mathbf{x}) \mathbf{I}\big)
\\
\label{eqn:tvae2}
    \mathbf{t} = \frac{\mathbf{z} - \mu}{\sqrt{\mathbf{W} \mathbf{u}}} \hspace{8mm}  p_{\theta}(\mathbf{x}| g_{\theta}(\mathbf{t})) =  p_{\theta}(\mathbf{x}| g_{\theta}(\mathbf{z}, \mathbf{u}))
\end{gather}
The parameters $\theta, \phi, \gamma$ and $\mu$ are then optimized to maximize the likelihood of the data through the Evidence Lower Bound (ELBO):
% \fontsize{9.5}{10}
\begin{equation}
    \label{eqn:elbo}
    \mathbb{E}_{q_{\phi}(\mathbf{z}|\mathbf{x})q_{\gamma}(\mathbf{u}|\mathbf{x})}
    \left(\log p_{\theta}(\mathbf{x}|g_{\theta}(\mathbf{t})) - D_{KL}[q_{\phi}(\mathbf{z}|\mathbf{x}) || p_{\mathbf{Z}}(\mathbf{z})] - D_{KL}[q_{\gamma}(\mathbf{u}|\mathbf{x}) || p_{\mathbf{U}}(\mathbf{u})]\right)
\end{equation}
% \normalsize
% \vspace{-3mm}

\newpage
\section{Methods}
% In this section we explain the datasets, evaluation metrics, and model architectures used to validate the Topographic VAE's emergent domain selectivity, as well as our reimplementation of the TDANN.

\paragraph{Evaluation}
Following prior computational work \cite{TDANN, VTCSOM} and fMRI studies \cite{Aparicio12729}, we use Cohen's $\mathit{d}$ metric \cite{cohen_1988, Sawilowsky_2009}, a measure of standardized difference of two means, as our selectivity metric. Given the means $\bar{m}_1$ \& $\bar{m}_2$ and standard deviations $\sigma_1$ \& $\sigma_2$ of two sets of data, the $\mathit{d}$ metric is given as: 
\begin{equation}
    \mathit{d} = \frac{\bar{m}_1 - \bar{m}_2}{\sqrt{\frac{1}{2}\left(\sigma^2_1 + \sigma^2_2\right)}}
\end{equation}
% \normalsize
This value is unitless and can be seen as expressing the difference between two means in terms of units of `pooled variability'. In this work, the mean $\bar{m}_1$ corresponds to the mean activation of a single neuron computed across an entire dataset of class-specific target images (e.g. faces), while $\bar{m}_2$ is the mean activation of the same neuron across a dataset of control images which do not contain this class.

\paragraph{Datasets}
The dataset used for training both the TDANN and TVAE is a composition of the  ImageNet 2012 \cite{ILSVRC15} and Labeled Faces in the Wild (LFW) datasets \cite{LFWTech}, following \citet{TDANN}. The TDANN was trained to classify the 1000 distinct image classes from ImageNet, plus one generic face class encompassing all of LFW. The TVAE used no such class labels. To measure the category selectivity of the models, the primary test face dataset used in Figures \ref{fig:selectivity}, \ref{fig:multiclass}, \& \ref{fig:hierarchy} was a $\sim$25,000 image subset of VGGface2 \cite{cao2018vggface2}. The control `object' dataset for Figures \ref{fig:selectivity} \& \ref{fig:hierarchy} was composed of 25,000 images from the validation set of ImageNet. To measure selectivity to body parts and places in Figure \ref{fig:multiclass}, we created a `body' dataset composed of headless body images \cite{body} and hands \cite{afifi201911kHands}, and used the Place365 dataset \cite{zhou2017places} for places. In Figure \ref{fig:multiclass}, the `control' set used for each class was defined to be the compliment of the test set, i.e. all other datasets besides the target category of interest.

% \subsection{Models}
\paragraph{Models} All models are trained on top of features extracted by the final convolutional layer of a pre-trained Alexnet model \cite{alexnet, pytorch}. The Alexnet architecture was chosen to match the setup from \citet{TDANN} and \citet{VTCSOM}, and has further been shown to have remarkable similarities to hierarchical processing in the human visual stream \cite{yamins2016, alexnet1, alexnet2}. For the TVAE, we randomly initialize and train a single linear layer encoder and decoder with 4096 output neurons, arranged in a 64x64 grid with circular boundary conditions to avoid edge effects. For the TDANN, we randomly initialize and train all three fully connected layers of Alexnet, imposing the spatial correlation loss over both `FC6' and `FC7'. In the following, all selectivity maps are displayed for `FC6', following \citet{TDANN}. All hyperparameter and training details can be found in Section \ref{appendix:model_details}.% for all hyperparameters and training details.

\section{Experiments}
In the following, we explore the category-selectivity of top-level neurons trained with the Topographic VAE framework on realistic images. We observe that neurons do indeed become category-selective, and that selective neurons tend to group together to form localized category-selective regions for a variety of domains including faces, bodies, and places. We compare these results with a non-topographic baseline (pre-trained Alexnet), and a re-implementation of the TDANN \cite{TDANN}, observing qualitatively similar results. Additionally, following \citet{VTCSOM}, we plot selectivity maps to more abstract concepts (such as animacy and real-world size), and observe that such maps overlap in an intuitive manner, suggesting the existence of a nested spatial hierarchy of categories. 

\begin{figure}[h]
\vspace{-2mm}
\centering
\includegraphics[width=1.0\linewidth]{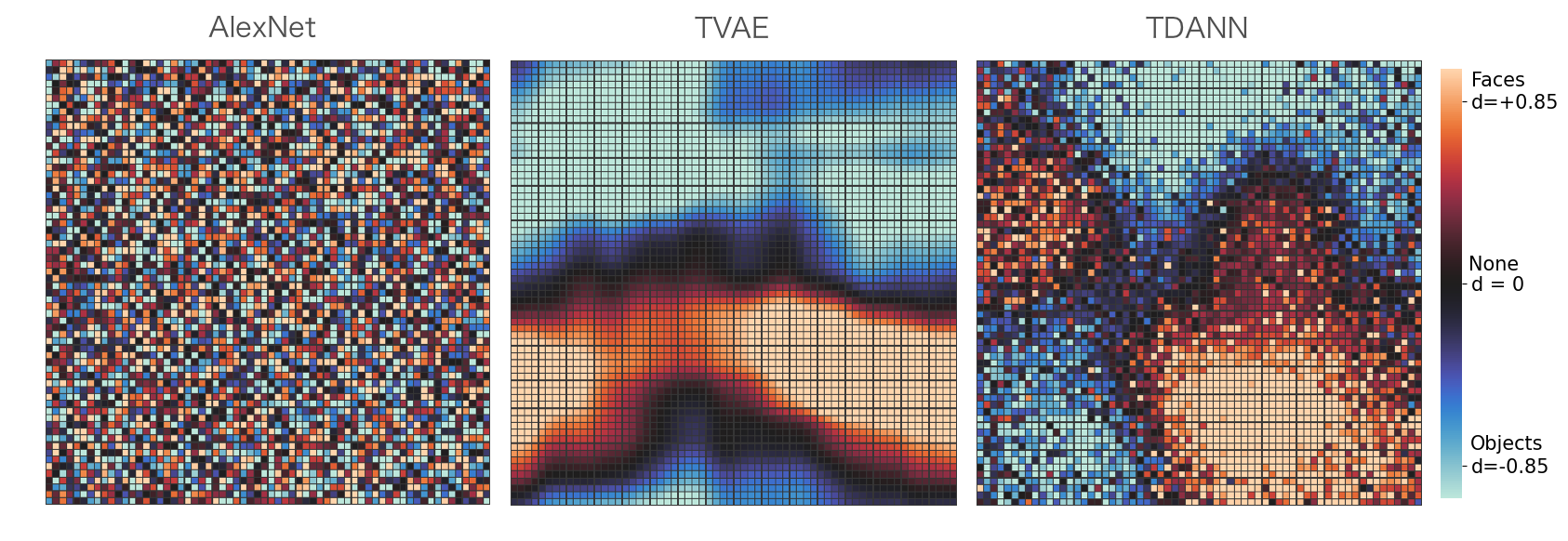}
\vspace{-4mm}
\caption{Face vs. Object selectivity for a non-topographic baseline, Topographic VAE, and TDANN. We see the TVAE has an emergent face cluster qualitatively similar to that of the TDANN.} 
\label{fig:selectivity}
\end{figure}

\paragraph{Localized Category-Selectivity}
In Figure \ref{fig:selectivity}, we plot the continuous value of Cohen's $\mathit{d}$ metric for all neurons as arranged in a 2-d grid. The baseline (left) shows the first fully connected layer (FC6) of a pre-trained Alexnet architecture. As expected, the neurons of this model have no defined spatial  organization and thus result in a random selectivity map. We note the existence of class-selective neurons is not guaranteed, but their appearance here is in-line with observations from prior work \cite{TDANN, selective1, selective2}. Secondly, we compare our TVAE model (middle) with our re-implementation of the TDANN \cite{TDANN} (right). We observe that both models demonstrate the emergence of face-selective clusters of comparable size and density.
We see that the TVAE framework appears to yield smoother topographic maps, perhaps due to the unified objective function and unsupervised learning rule when compared with the competing supervised classification loss and wiring cost regularization of the TDANN. To validate the robustness and significance of these category selective regions, in Section \ref{appendix:additional_results} of the appendix we plot selectivity maps across four different test face datasets and four random initalizations, observing qualitatively similar clusters all settings.

\vspace{-3mm}
\paragraph{Face, Body \& Place Clusters} 
\begin{wrapfigure}{r}{0.38\linewidth}
\centering
\vspace{-4mm}
\includegraphics[width=1.0\linewidth]{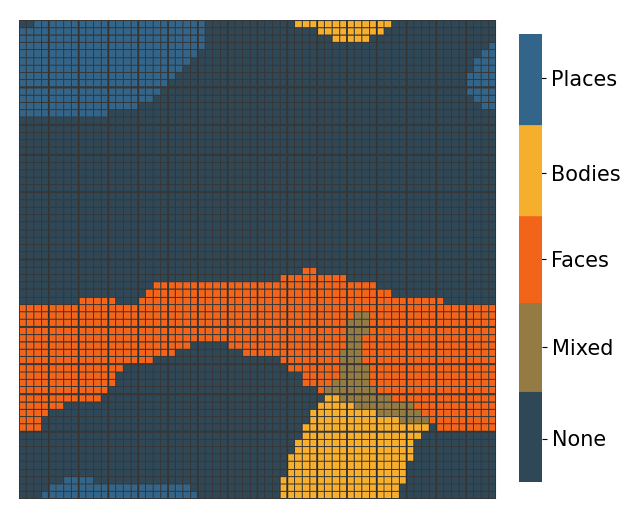}
\vspace{-5mm}
\caption{TVAE selectivity $\mathit{d}\geq 0.85$}
\label{fig:multiclass}
\vspace{-7mm}
\end{wrapfigure}
Next, in Figure \ref{fig:multiclass}, we plot the simultaneous selectivity of neurons in our TVAE model to multiple classes including faces, bodies, and places. To create a map of multi-class selectivity, we follow prior work and threshold the $\mathit{d}$ metric at $0.85$, considered a `strong effect' \cite{Sawilowsky_2009} and computed to be to be equivalent to a threshold of $0.65$ for noisy neural recordings in monkeys \cite{TDANN}. In the plot we observe an overlap of neurons with selectivity to faces and bodies, as seen in prior computational work \cite{TDANN} and fMRI studies \cite{Pinsk6996, facebody}. In Figure \ref{fig:all_robustness} of Section \ref{appendix:additional_results}, we see that the size and relative placement of these clusters is again consistent across multiple random initalizations.

% \vspace{2mm}
\paragraph{Impact of Topography on Model Performance}
To measure the impact of the imposed topographic organization on the above models, and ensure the learned representations are not degenerate, we compare the model performance of the TDANN and TVAE with their respective non-topographic counterparts. Although the models in this study were not tuned to maximize such performance, we observe that both topographic models perform similarly to their non-topographic counterparts. Specifically the TDANN achieves 40.5\% top-1 accuracy on the Imagenet validation set (+ 1 face class) versus the 45.5\% top-1 accuracy of an identically trained model without spatial correlation loss. Similarly, a baseline VAE of the same architecture as the TVAE achieves roughly 3.4 bits per dimension (BPD) while the Topographic VAE achieves roughly 3.6 BPD in the same number of iterations. These results appear consistent with the intuition that topographic organization does not prevent learning, but rather acts as an inductive bias on the model, regularizing its performance. In future work we hope to quantify this regularization effect more precisely and determine in which situations it may be beneficial for generalization or computational efficiency.

% \vspace{2mm}

\begin{figure}[h!]
\centering
% \vspace{-3mm}
\includegraphics[width=1.0\linewidth]{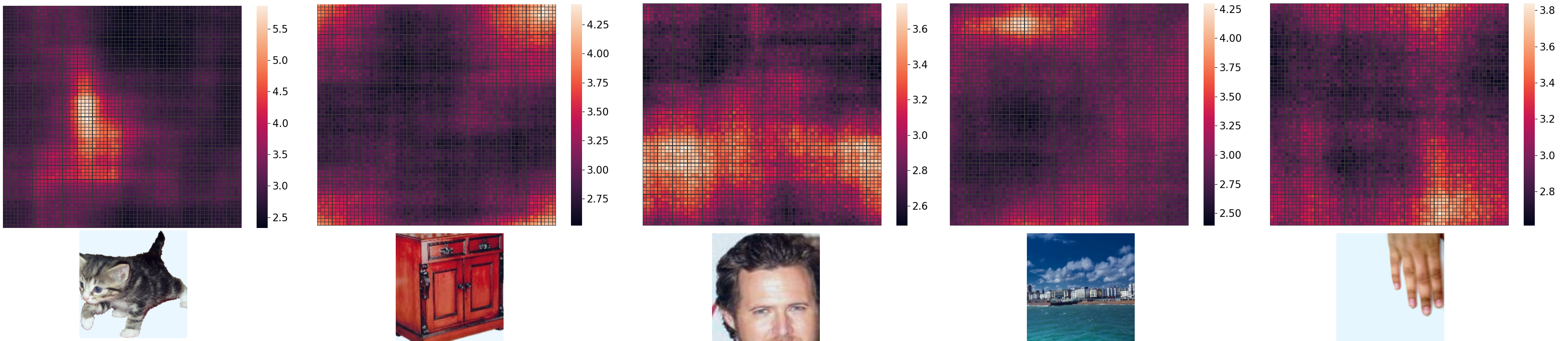}
\vspace{-1mm}
\caption{
Activations for single images. Left to Right: Animate, Inanimate, Faces, Places, \& Hands \vspace{-5mm}} 
\label{fig:single_img_maps}
\end{figure}

\paragraph{Locally Distributed Activations}
To understand better how exactly individual images are represented by the TVAE, we present the activation maps corresponding to a single image from an array of classes in Figure \ref{fig:single_img_maps}. We see the representation of each image is still distributed, but most strongly activates in the associated category-selective region.

\begin{wrapfigure}{r}{0.5\linewidth}
\centering
\vspace{-17mm}
\includegraphics[width=1.0\linewidth]{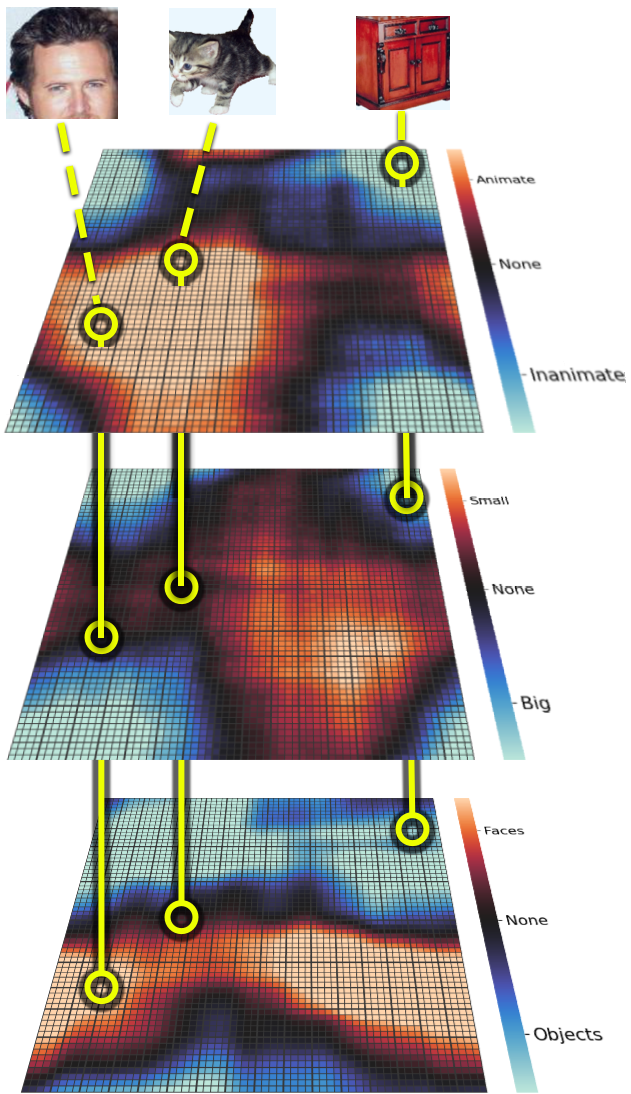}
\vspace{-2mm}
\caption{Selectivity maps for abstract categories: Animate vs. Inanimate (top), Small vs. Big (middle), and Faces vs. Objects (bottom). We highlight the maximum activating neurons for the individual images from Figure \ref{fig:single_img_maps} across all maps, demonstrating their place in the proposed nested spatial hierarchy.} 
\label{fig:hierarchy}
\vspace{-13mm}
\end{wrapfigure}

% \vspace{-18mm}
\paragraph{Nested Spatial Hierarchy of Categories}
Following \citet{VTCSOM} we additionally measure the selectivity maps of our TVAE model with respect to more abstract categories such as animacy and real-world object size, obtaining such datasets from the Konkle lab database \cite{size, Konkle10235}. Specifically, Figure \ref{fig:hierarchy} shows Cohen's $\mathit{d}$ maps (from $-1$ to $+1$) for animate versus inaminate objects (top), and for big versus small objects (middle), overlayed on the face versus object map (bottom). At the largest scale, we observe an intuitive overlap of spatial maps, specifically inanimate objects, large objects, and the place cluster from Figure \ref{fig:multiclass} all overlap in the top left and right corners of the map. We additionally highlight the maximum activating neurons for three separate input images. We see the image of a red dresser activates a region which is simultaneously selective to places, large, and inanimate objects, echoing the nested spatial hierarchies thought by \citet{stanfordsurvey} to exist in the brain. 
In Section \ref{appendix:additional_results}, we again see that such a hierarchy appears consistently across four random initalizations.

\vspace{8mm}

\vspace{-8mm}
\section{Discussion}
% \vspace{-3mm}
In this work we demonstrate the ability of topographic generative models, namely Topographic Variational Autoencoders, to model the emergence of category-selective cortical areas as well as more abstract spatial category hierarchies. We see the model agrees qualitatively with prior work and observations from neuroscience while being founded on a single information theoretic principle. 

We note that this study is inherently preliminary and is  limited by both the small size of the models used, as well as the feature extraction by a pre-trained convolutional model. It is possible that class-level features and even hierarchical organization are already partially present in some form in the 9216-dimensional feature vectors used as input, and thus it is unclear how much feature extraction the TVAE model is itself learning. Nevertheless, we highlight that there is nothing fundamentally limiting the TVAE framework from extending to train full deep convolutional networks end-to-end. This is in contrast to the existing related methods which either require a supplementary learning signal to guide feature extraction \cite{TDANN}, or do not scale to high dimensional inputs \cite{VTCSOM}.

In future work, we intend to explore hierarchical extensions of the TVAE, modeling topographic organization of features at multiple levels of the visual processing pipeline while simultaneously training  directly on raw pixel inputs. Such a model would validate the idea of end-to-end unsupervised category-selectivity while simultaneously providing a learned decoder from latent space to image space, opening new avenues for experimentation.

\begin{ack}
We would like to thank the reviewers for providing helpful constructive feedback, and the organizers of the workshop for their service. Additionally, we thank the creators of Weight \& Biases \cite{wandb} and PyTorch \cite{pytorch}. Finally, we thank the Bosch Center for Artificial Intelligence for funding.
\end{ack}

\bibliography{svrhm_2021}

\begin{thebibliography}{65}
\providecommand{\natexlab}[1]{#1}
\providecommand{\url}[1]{\texttt{#1}}
\expandafter\ifx\csname urlstyle\endcsname\relax
  \providecommand{\doi}[1]{doi: #1}\else
  \providecommand{\doi}{doi: \begingroup \urlstyle{rm}\Url}\fi

\bibitem[TEN(2006)]{TENENBAUM2006309}
Theory-based bayesian models of inductive learning and reasoning.
\newblock \emph{Trends in Cognitive Sciences}, 10\penalty0 (7):\penalty0
  309--318, 2006.
\newblock ISSN 1364-6613.
\newblock \doi{https://doi.org/10.1016/j.tics.2006.05.009}.
\newblock URL
  \url{https://www.sciencedirect.com/science/article/pii/S1364661306001343}.
\newblock Special issue: Probabilistic models of cognition.

\bibitem[sel(2019)]{selective2}
Convergent evolution of face spaces across human face-selective neuronal groups
  and deep convolutional networks.
\newblock \emph{Nature Communications}, 10\penalty0 (1):\penalty0 4934, 2019.
\newblock \doi{10.1038/s41467-019-12623-6}.
\newblock URL \url{https://doi.org/10.1038/s41467-019-12623-6}.

\bibitem[Afifi(2019)]{afifi201911kHands}
Mahmoud Afifi.
\newblock 11k hands: gender recognition and biometric identification using a
  large dataset of hand images.
\newblock \emph{Multimedia Tools and Applications}, 2019.
\newblock \doi{10.1007/s11042-019-7424-8}.
\newblock URL \url{https://doi.org/10.1007/s11042-019-7424-8}.

\bibitem[Aparicio et~al.(2016)Aparicio, Issa, and DiCarlo]{Aparicio12729}
Paul~L. Aparicio, Elias~B. Issa, and James~J. DiCarlo.
\newblock Neurophysiological organization of the middle face patch in macaque
  inferior temporal cortex.
\newblock \emph{Journal of Neuroscience}, 36\penalty0 (50):\penalty0
  12729--12745, 2016.
\newblock ISSN 0270-6474.
\newblock \doi{10.1523/JNEUROSCI.0237-16.2016}.
\newblock URL \url{https://www.jneurosci.org/content/36/50/12729}.

\bibitem[Bakhtiari et~al.(2021)Bakhtiari, Mineault, Lillicrap, Pack, and
  Richards]{functional_spec}
Shahab Bakhtiari, Patrick Mineault, Tim Lillicrap, Christopher~C. Pack, and
  Blake~A. Richards.
\newblock The functional specialization of visual cortex emerges from training
  parallel pathways with self-supervised predictive learning.
\newblock \emph{bioRxiv}, 2021.
\newblock \doi{10.1101/2021.06.18.448989}.
\newblock URL
  \url{https://www.biorxiv.org/content/early/2021/06/24/2021.06.18.448989}.

\bibitem[Ball{\'e} et~al.(2016)Ball{\'e}, Laparra, and Simoncelli]{GDN}
Johannes Ball{\'e}, Valero Laparra, and {Eero P.} Simoncelli.
\newblock Density modeling of images using a generalized normalization
  transformation.
\newblock January 2016.
\newblock 4th International Conference on Learning Representations, ICLR 2016.

\bibitem[Barlow et~al.(1961)]{barlow1961possible}
Horace~B Barlow et~al.
\newblock Possible principles underlying the transformation of sensory
  messages.
\newblock \emph{Sensory communication}, 1\penalty0 (01), 1961.

\bibitem[Bell \& Sejnowski(1995)Bell and Sejnowski]{SejnowskiICA}
Anthony~J. Bell and Terrence~J. Sejnowski.
\newblock {An Information-Maximization Approach to Blind Separation and Blind
  Deconvolution}.
\newblock \emph{Neural Computation}, 7\penalty0 (6):\penalty0 1129--1159, 11
  1995.
\newblock ISSN 0899-7667.
\newblock \doi{10.1162/neco.1995.7.6.1129}.
\newblock URL \url{https://doi.org/10.1162/neco.1995.7.6.1129}.

\bibitem[Biewald(2020)]{wandb}
Lukas Biewald.
\newblock Experiment tracking with weights and biases, 2020.
\newblock URL \url{https://www.wandb.com/}.
\newblock Software available from wandb.com.

\bibitem[Blauch et~al.(2021)Blauch, Behrmann, and Plaut]{Blauch2021}
Nicholas~M. Blauch, Marlene Behrmann, and David~C. Plaut.
\newblock A connectivity-constrained computational account of topographic
  organization in primate high-level visual cortex.
\newblock \emph{bioRxiv}, 2021.
\newblock \doi{10.1101/2021.05.29.446297}.
\newblock URL
  \url{https://www.biorxiv.org/content/early/2021/07/12/2021.05.29.446297}.

\bibitem[Cao et~al.(2018)Cao, Shen, Xie, Parkhi, and
  Zisserman]{cao2018vggface2}
Qiong Cao, Li~Shen, Weidi Xie, Omkar~M Parkhi, and Andrew Zisserman.
\newblock Vggface2: A dataset for recognising faces across pose and age.
\newblock In \emph{2018 13th IEEE international conference on automatic face \&
  gesture recognition (FG 2018)}, pp.\  67--74. IEEE, 2018.

\bibitem[Caspers et~al.(2013)Caspers, Palomero-Gallagher, Caspers, Schleicher,
  Amunts, and Zilles]{receptors}
Julian Caspers, Nicola Palomero-Gallagher, Svenja Caspers, Axel Schleicher,
  Katrin Amunts, and Karl Zilles.
\newblock Receptor architecture of visual areas in the face and word-form
  recognition region of the posterior fusiform gyrus.
\newblock \emph{Brain structure \& function}, 220, 10 2013.
\newblock \doi{10.1007/s00429-013-0646-z}.

\bibitem[Cichy et~al.(2016)Cichy, Khosla, Pantazis, Torralba, and
  Oliva]{alexnet1}
Radoslaw~Martin Cichy, Aditya Khosla, Dimitrios Pantazis, Antonio Torralba, and
  Aude Oliva.
\newblock Comparison of deep neural networks to spatio-temporal cortical
  dynamics of human visual object recognition reveals hierarchical
  correspondence.
\newblock \emph{Scientific Reports}, 6\penalty0 (1):\penalty0 27755, 2016.
\newblock \doi{10.1038/srep27755}.
\newblock URL \url{https://doi.org/10.1038/srep27755}.

\bibitem[Cohen(1988)]{cohen_1988}
Jack Cohen.
\newblock \emph{Statistical Power Analysis for the behavioral sciences}.
\newblock L. Erlbaum Associates, 1988.

\bibitem[Comon(1994)]{comon1994independent}
Pierre Comon.
\newblock Independent component analysis, a new concept?
\newblock \emph{Signal processing}, 36\penalty0 (3):\penalty0 287--314, 1994.

\bibitem[Dobs et~al.(2021)Dobs, Martinez, Kell, and Kanwisher]{Dobs2021}
Katharina Dobs, Julio Martinez, Alexander~J.E. Kell, and Nancy Kanwisher.
\newblock Brain-like functional specialization emerges spontaneously in deep
  neural networks.
\newblock \emph{bioRxiv}, 2021.
\newblock \doi{10.1101/2021.07.05.451192}.
\newblock URL
  \url{https://www.biorxiv.org/content/early/2021/07/06/2021.07.05.451192}.

\bibitem[Epstein \& Kanwisher(1998)Epstein and Kanwisher]{ppa}
Russell~A. Epstein and Nancy Kanwisher.
\newblock A cortical representation of the local visual environment.
\newblock \emph{Nature}, 392:\penalty0 598--601, 1998.

\bibitem[Essen(1997)]{tension}
David C.~Van Essen.
\newblock A tension-based theory of morphogenesis and compact wiring in the
  central nervous system.
\newblock \emph{Nature}, 385\penalty0 (6614):\penalty0 313--318, 1997.
\newblock \doi{10.1038/385313a0}.
\newblock URL \url{https://doi.org/10.1038/385313a0}.

\bibitem[Grill-Spector \& Weiner(2014)Grill-Spector and Weiner]{stanfordsurvey}
Kalanit Grill-Spector and Kevin~S. Weiner.
\newblock The functional architecture of the ventral temporal cortex and its
  role in categorization.
\newblock \emph{Nature Reviews Neuroscience}, 15\penalty0 (8):\penalty0
  536--548, 2014.
\newblock \doi{10.1038/nrn3747}.
\newblock URL \url{https://doi.org/10.1038/nrn3747}.

\bibitem[G{\"u}{\c c}l{\"u} \& van Gerven(2015)G{\"u}{\c c}l{\"u} and van
  Gerven]{alexnet2}
Umut G{\"u}{\c c}l{\"u} and Marcel A.~J. van Gerven.
\newblock Deep neural networks reveal a gradient in the complexity of neural
  representations across the ventral stream.
\newblock \emph{Journal of Neuroscience}, 35\penalty0 (27):\penalty0
  10005--10014, 2015.
\newblock ISSN 0270-6474.
\newblock \doi{10.1523/JNEUROSCI.5023-14.2015}.
\newblock URL \url{https://www.jneurosci.org/content/35/27/10005}.

\bibitem[Haxby et~al.(2001)Haxby, Gobbini, Furey, Ishai, Schouten, and
  Pietrini]{hierarchy}
James Haxby, Maria Gobbini, Maura Furey, Alumit Ishai, Jennifer Schouten, and
  Pietro Pietrini.
\newblock Distributed and overlapping representations of faces and objects in
  ventral temporal cortex.
\newblock \emph{Science (New York, N.Y.)}, 293:\penalty0 2425--30, 10 2001.
\newblock \doi{10.1126/science.1063736}.

\bibitem[Haxby et~al.(2011)Haxby, Guntupalli, Connolly, Halchenko, Conroy,
  Gobbini, Hanke, and Ramadge]{animacy}
James Haxby, Jyothi~Swaroop Guntupalli, Andrew Connolly, Yaroslav Halchenko,
  Bryan Conroy, Maria Gobbini, Michael Hanke, and Peter Ramadge.
\newblock A common, high-dimensional model of the representational space in
  human ventral temporal cortex.
\newblock \emph{Neuron}, 72:\penalty0 404--16, 10 2011.
\newblock \doi{10.1016/j.neuron.2011.08.026}.

\bibitem[Huang et~al.(2007)Huang, Ramesh, Berg, and Learned-Miller]{LFWTech}
Gary~B. Huang, Manu Ramesh, Tamara Berg, and Erik Learned-Miller.
\newblock Labeled faces in the wild: A database for studying face recognition
  in unconstrained environments.
\newblock Technical Report 07-49, University of Massachusetts, Amherst, October
  2007.

\bibitem[Hyv{\"a}rinen(1998)]{hyvarinen1998independent}
Aapo Hyv{\"a}rinen.
\newblock
  \href{https://www.sciencedirect.com/science/article/pii/S0925231298000496?casa_token=R59e056ns9gAAAAA:QfbGXxJW8hcjScbkQ2wF5sSCKJ3wWNH9U6QbHPyMMBVAAaC9xwKEA2gQkSOSwlBanUkob3xOAsE}{Independent
  component analysis in the presence of gaussian noise by maximizing joint
  likelihood}.
\newblock \emph{Neurocomputing}, 22\penalty0 (1-3):\penalty0 49--67, 1998.

\bibitem[Hyv{\"a}rinen \& Oja(2000)Hyv{\"a}rinen and
  Oja]{hyvarinen2000independent}
Aapo Hyv{\"a}rinen and Erkki Oja.
\newblock
  \href{https://www.cs.helsinki.fi/u/ahyvarin/papers/bookfinal_ICA.pdf}{Independent
  component analysis: algorithms and applications}.
\newblock \emph{Neural networks}, 13\penalty0 (4-5):\penalty0 411--430, 2000.

\bibitem[Hyv{\"a}rinen et~al.(2001)Hyv{\"a}rinen, Hoyer, and
  Inki]{hyvarinen2001topographic}
Aapo Hyv{\"a}rinen, Patrik~O Hoyer, and Mika Inki.
\newblock
  \href{https://www.cs.helsinki.fi/u/ahyvarin/papers/NC01_TICA.pdf}{Topographic
  independent component analysis}.
\newblock \emph{Neural computation}, 13\penalty0 (7):\penalty0 1527--1558,
  2001.

\bibitem[Hyv{\"a}rinen et~al.(2009)Hyv{\"a}rinen, Hurri, and
  Hoyer]{hyvarinen2009natural}
Aapo Hyv{\"a}rinen, Jarmo Hurri, and Patrick~O Hoyer.
\newblock \emph{Natural image statistics: A probabilistic approach to early
  computational vision.}, volume~39.
\newblock Springer Science \& Business Media, 2009.

\bibitem[Hyvärinen \& Hoyer(2001)Hyvärinen and Hoyer]{HYVARINEN20012413}
Aapo Hyvärinen and Patrik~O. Hoyer.
\newblock A two-layer sparse coding model learns simple and complex cell
  receptive fields and topography from natural images.
\newblock \emph{Vision Research}, 41\penalty0 (18):\penalty0 2413--2423, 2001.
\newblock ISSN 0042-6989.
\newblock \doi{https://doi.org/10.1016/S0042-6989(01)00114-6}.
\newblock URL
  \url{https://www.sciencedirect.com/science/article/pii/S0042698901001146}.

\bibitem[Kanwisher et~al.(1997)Kanwisher, McDermott, and Chun]{ffa}
Nancy Kanwisher, Josh McDermott, and Marvin~M. Chun.
\newblock The fusiform face area: A module in human extrastriate cortex
  specialized for face perception.
\newblock \emph{Journal of Neuroscience}, 17\penalty0 (11):\penalty0
  4302--4311, 1997.
\newblock ISSN 0270-6474.
\newblock \doi{10.1523/JNEUROSCI.17-11-04302.1997}.
\newblock URL \url{https://www.jneurosci.org/content/17/11/4302}.

\bibitem[Keller \& Welling(2021{\natexlab{a}})Keller and Welling]{PCTVAE}
T.~Anderson Keller and Max Welling.
\newblock Predictive coding with topographic variational autoencoders.
\newblock In \emph{Proceedings of the IEEE/CVF International Conference on
  Computer Vision (ICCV) Workshops}, pp.\  1086--1091, October
  2021{\natexlab{a}}.

\bibitem[Keller \& Welling(2021{\natexlab{b}})Keller and Welling]{TVAE}
T.~Anderson Keller and Max Welling.
\newblock Topographic vaes learn equivariant capsules.
\newblock In \emph{Advances in Neural Information Processing Systems 34},
  December 2021{\natexlab{b}}.

\bibitem[Kingma \& Welling(2014)Kingma and Welling]{kingma2013auto}
Diederik~P. Kingma and Max Welling.
\newblock Auto-encoding variational bayes.
\newblock In Yoshua Bengio and Yann LeCun (eds.), \emph{2nd International
  Conference on Learning Representations, {ICLR} 2014, Banff, AB, Canada, April
  14-16, 2014, Conference Track Proceedings}, 2014.
\newblock URL \url{http://arxiv.org/abs/1312.6114}.

\bibitem[Konkle \& Alvarez(2021)Konkle and Alvarez]{konkle_selfsup}
Talia Konkle and George~A. Alvarez.
\newblock Beyond category-supervision: Computational support for domain-general
  pressures guiding human visual system representation.
\newblock \emph{bioRxiv}, 2021.
\newblock \doi{10.1101/2020.06.15.153247}.
\newblock URL
  \url{https://www.biorxiv.org/content/early/2021/10/19/2020.06.15.153247}.

\bibitem[Konkle \& Caramazza(2013)Konkle and Caramazza]{Konkle10235}
Talia Konkle and Alfonso Caramazza.
\newblock Tripartite organization of the ventral stream by animacy and object
  size.
\newblock \emph{Journal of Neuroscience}, 33\penalty0 (25):\penalty0
  10235--10242, 2013.
\newblock ISSN 0270-6474.
\newblock \doi{10.1523/JNEUROSCI.0983-13.2013}.
\newblock URL \url{https://www.jneurosci.org/content/33/25/10235}.

\bibitem[Konkle \& Oliva(2012)Konkle and Oliva]{size}
Talia Konkle and Aude Oliva.
\newblock A real-world size organization of object responses in
  occipitotemporal cortex.
\newblock \emph{Neuron}, 74:\penalty0 1114--24, 06 2012.
\newblock \doi{10.1016/j.neuron.2012.04.036}.

\bibitem[Koulakov \& Chklovskii(2001)Koulakov and
  Chklovskii]{koulakov2001orientation}
Alexei~A Koulakov and Dmitri~B Chklovskii.
\newblock Orientation preference patterns in mammalian visual cortex: a wire
  length minimization approach.
\newblock \emph{Neuron}, 29\penalty0 (2):\penalty0 519--527, 2001.

\bibitem[Krizhevsky et~al.(2012)Krizhevsky, Sutskever, and Hinton]{alexnet}
Alex Krizhevsky, Ilya Sutskever, and Geoffrey~E Hinton.
\newblock Imagenet classification with deep convolutional neural networks.
\newblock In F.~Pereira, C.~J.~C. Burges, L.~Bottou, and K.~Q. Weinberger
  (eds.), \emph{Advances in Neural Information Processing Systems}, volume~25.
  Curran Associates, Inc., 2012.
\newblock URL
  \url{https://proceedings.neurips.cc/paper/2012/file/c399862d3b9d6b76c8436e924a68c45b-Paper.pdf}.

\bibitem[Leavitt \& Morcos(2021)Leavitt and Morcos]{leavitt2020selectivity}
Matthew~L Leavitt and Ari~S. Morcos.
\newblock Selectivity considered harmful: evaluating the causal impact of class
  selectivity in {\{}dnn{\}}s.
\newblock In \emph{International Conference on Learning Representations}, 2021.
\newblock URL \url{https://openreview.net/forum?id=8nl0k08uMi}.

\bibitem[Lee et~al.(2020)Lee, Margalit, Jozwik, Cohen, Kanwisher, Yamins, and
  DiCarlo]{TDANN}
Hyodong Lee, Eshed Margalit, Kamila~M. Jozwik, Michael~A. Cohen, Nancy
  Kanwisher, Daniel L.~K. Yamins, and James~J. DiCarlo.
\newblock Topographic deep artificial neural networks reproduce the hallmarks
  of the primate inferior temporal cortex face processing network.
\newblock \emph{bioRxiv}, 2020.
\newblock \doi{10.1101/2020.07.09.185116}.
\newblock URL
  \url{https://www.biorxiv.org/content/early/2020/07/10/2020.07.09.185116}.

\bibitem[Liu et~al.(2015)Liu, Luo, Wang, and Tang]{liu2015faceattributes}
Ziwei Liu, Ping Luo, Xiaogang Wang, and Xiaoou Tang.
\newblock Deep learning face attributes in the wild.
\newblock In \emph{Proceedings of International Conference on Computer Vision
  (ICCV)}, December 2015.

\bibitem[Lyu \& Simoncelli(2008)Lyu and Simoncelli]{simoncelli}
Siwei Lyu and Eero~P. Simoncelli.
\newblock Nonlinear image representation using divisive normalization.
\newblock In \emph{2008 IEEE Conference on Computer Vision and Pattern
  Recognition}, pp.\  1--8, 2008.
\newblock \doi{10.1109/CVPR.2008.4587821}.

\bibitem[Ma \& Zhang(2008)Ma and Zhang]{ma2008overcomplete}
Libo Ma and Liqing Zhang.
\newblock
  \href{https://reader.elsevier.com/reader/sd/pii/S0925231208000957?token=FEFABF17DBC9D13F8F221FA7EC2F1AE1F07AA22B11FFEE0E90383CF39C696F0DADBE29276AC1CD7FF6F2F13E63846C99}{Overcomplete
  topographic independent component analysis}.
\newblock \emph{Neurocomputing}, 71\penalty0 (10-12):\penalty0 2217--2223,
  2008.

\bibitem[Moro et~al.(2008)Moro, Urgesi, Pernigo, Lanteri, Pazzaglia, and
  Aglioti]{lesionPPA}
Valentina Moro, Cosimo Urgesi, Simone Pernigo, Paola Lanteri, Mariella
  Pazzaglia, and Salvatore Aglioti.
\newblock The neural basis of body form and body action agnosia.
\newblock \emph{Neuron}, 60:\penalty0 235--46, 11 2008.
\newblock \doi{10.1016/j.neuron.2008.09.022}.

\bibitem[Nasr et~al.(2011)Nasr, Liu, Devaney, Yue, Rajimehr, Ungerleider, and
  Tootell]{monkeyppa}
Shahin Nasr, Ning Liu, Kathryn~J. Devaney, Xiaomin Yue, Reza Rajimehr,
  Leslie~G. Ungerleider, and Roger B.~H. Tootell.
\newblock Scene-selective cortical regions in human and nonhuman primates.
\newblock \emph{Journal of Neuroscience}, 31\penalty0 (39):\penalty0
  13771--13785, 2011.
\newblock ISSN 0270-6474.
\newblock \doi{10.1523/JNEUROSCI.2792-11.2011}.
\newblock URL \url{https://www.jneurosci.org/content/31/39/13771}.

\bibitem[Olshausen \& Field(1997)Olshausen and Field]{olshausen1997sparse}
Bruno~A Olshausen and David~J Field.
\newblock \href{shorturl.at/byGPV}{Sparse coding with an overcomplete basis
  set: A strategy employed by V1?}
\newblock \emph{Vision research}, 37\penalty0 (23):\penalty0 3311--3325, 1997.

\bibitem[Osindero et~al.(2006)Osindero, Welling, and Hinton]{Osindero2006}
Simon Osindero, Max Welling, and Geoffrey~E. Hinton.
\newblock {Topographic Product Models Applied to Natural Scene Statistics}.
\newblock \emph{Neural Computation}, 18\penalty0 (2):\penalty0 381--414, 02
  2006.
\newblock ISSN 0899-7667.
\newblock \doi{10.1162/089976606775093936}.
\newblock URL \url{https://doi.org/10.1162/089976606775093936}.

\bibitem[Paszke et~al.(2019)Paszke, Gross, Massa, Lerer, Bradbury, Chanan,
  Killeen, Lin, Gimelshein, Antiga, Desmaison, Kopf, Yang, DeVito, Raison,
  Tejani, Chilamkurthy, Steiner, Fang, Bai, and Chintala]{pytorch}
Adam Paszke, Sam Gross, Francisco Massa, Adam Lerer, James Bradbury, Gregory
  Chanan, Trevor Killeen, Zeming Lin, Natalia Gimelshein, Luca Antiga, Alban
  Desmaison, Andreas Kopf, Edward Yang, Zachary DeVito, Martin Raison, Alykhan
  Tejani, Sasank Chilamkurthy, Benoit Steiner, Lu~Fang, Junjie Bai, and Soumith
  Chintala.
\newblock Pytorch: An imperative style, high-performance deep learning library.
\newblock In \emph{Advances in Neural Information Processing Systems 32}, pp.\
  8024--8035. 2019.

\bibitem[Peelen \& Downing(2005)Peelen and Downing]{eba}
Marius~V. Peelen and Paul~E. Downing.
\newblock Selectivity for the human body in the fusiform gyrus.
\newblock \emph{Journal of Neurophysiology}, 93\penalty0 (1):\penalty0
  603--608, 2005.
\newblock \doi{10.1152/jn.00513.2004}.
\newblock URL \url{https://doi.org/10.1152/jn.00513.2004}.
\newblock PMID: 15295012.

\bibitem[Pinsk et~al.(2005{\natexlab{a}})Pinsk, DeSimone, Moore, Gross, and
  Kastner]{Pinsk6996}
Mark~A. Pinsk, Kevin DeSimone, Tirin Moore, Charles~G. Gross, and Sabine
  Kastner.
\newblock Representations of faces and body parts in macaque temporal cortex: A
  functional mri study.
\newblock \emph{Proceedings of the National Academy of Sciences}, 102\penalty0
  (19):\penalty0 6996--7001, 2005{\natexlab{a}}.
\newblock ISSN 0027-8424.
\newblock \doi{10.1073/pnas.0502605102}.
\newblock URL \url{https://www.pnas.org/content/102/19/6996}.

\bibitem[Pinsk et~al.(2005{\natexlab{b}})Pinsk, DeSimone, Moore, Gross, and
  Kastner]{monkeyfmri}
Mark~A. Pinsk, Kevin DeSimone, Tirin Moore, Charles~G. Gross, and Sabine
  Kastner.
\newblock Representations of faces and body parts in macaque temporal cortex: A
  functional mri study.
\newblock \emph{Proceedings of the National Academy of Sciences}, 102\penalty0
  (19):\penalty0 6996--7001, 2005{\natexlab{b}}.
\newblock ISSN 0027-8424.
\newblock \doi{10.1073/pnas.0502605102}.
\newblock URL \url{https://www.pnas.org/content/102/19/6996}.

\bibitem[Raman \& Hosoya(2020)Raman and Hosoya]{selective1}
Rajani Raman and Haruo Hosoya.
\newblock Convolutional neural networks explain tuning properties of anterior,
  but not middle, face-processing areas in macaque inferotemporal cortex.
\newblock \emph{Communications Biology}, 3\penalty0 (1):\penalty0 221, 2020.
\newblock \doi{10.1038/s42003-020-0945-x}.
\newblock URL \url{https://doi.org/10.1038/s42003-020-0945-x}.

\bibitem[Rezende et~al.(2014)Rezende, Mohamed, and
  Wierstra]{rezende2014stochastic}
Danilo~Jimenez Rezende, Shakir Mohamed, and Daan Wierstra.
\newblock Stochastic backpropagation and approximate inference in deep
  generative models.
\newblock In Eric~P. Xing and Tony Jebara (eds.), \emph{Proceedings of the 31st
  International Conference on Machine Learning}, volume~32 of \emph{Proceedings
  of Machine Learning Research}, pp.\  1278--1286, Bejing, China, 22--24 Jun
  2014. PMLR.
\newblock URL \url{https://proceedings.mlr.press/v32/rezende14.html}.

\bibitem[Russakovsky et~al.(2015)Russakovsky, Deng, Su, Krause, Satheesh, Ma,
  Huang, Karpathy, Khosla, Bernstein, Berg, and Fei-Fei]{ILSVRC15}
Olga Russakovsky, Jia Deng, Hao Su, Jonathan Krause, Sanjeev Satheesh, Sean Ma,
  Zhiheng Huang, Andrej Karpathy, Aditya Khosla, Michael Bernstein,
  Alexander~C. Berg, and Li~Fei-Fei.
\newblock {ImageNet Large Scale Visual Recognition Challenge}.
\newblock \emph{International Journal of Computer Vision (IJCV)}, 115\penalty0
  (3):\penalty0 211--252, 2015.
\newblock \doi{10.1007/s11263-015-0816-y}.

\bibitem[Sawilowsky(2009)]{Sawilowsky_2009}
Shlomo~S. Sawilowsky.
\newblock New effect size rules of thumb.
\newblock \emph{Journal of Modern Applied Statistical Methods}, 8\penalty0
  (2):\penalty0 597–599, Nov 2009.
\newblock \doi{10.22237/jmasm/1257035100}.
\newblock URL \url{http://dx.doi.org/10.22237/jmasm/1257035100}.

\bibitem[Saygin et~al.(2012)Saygin, Osher, Koldewyn, Reynolds, Gabrieli, and
  Saxe]{anatomy}
Zeynep~M. Saygin, David~E. Osher, Kami Koldewyn, Gretchen~O Reynolds, John
  D.~E. Gabrieli, and Rebecca Saxe.
\newblock Anatomical connectivity patterns predict face-selectivity in the
  fusiform gyrus.
\newblock \emph{Nature neuroscience}, 15:\penalty0 321 -- 327, 2012.

\bibitem[Tenenbaum(1999)]{Tenenbaum1998BayesianMO}
Joshua Tenenbaum.
\newblock Bayesian modeling of human concept learning.
\newblock In M.~Kearns, S.~Solla, and D.~Cohn (eds.), \emph{Advances in Neural
  Information Processing Systems}, volume~11. MIT Press, 1999.
\newblock URL
  \url{https://proceedings.neurips.cc/paper/1998/file/d010396ca8abf6ead8cacc2c2f2f26c7-Paper.pdf}.

\bibitem[{Tsao} et~al.(2006){Tsao}, {Freiwald}, {Tootell}, and
  {Livingstone}]{monkeyffa}
Doris~Y. {Tsao}, Winrich~A. {Freiwald}, Roger B.~H. {Tootell}, and Margaret~S.
  {Livingstone}.
\newblock A cortical region consisting entirely of face-selective cells.
\newblock \emph{Science}, 311\penalty0 (5761):\penalty0 670--674, 2006.

\bibitem[Weiner \& Grill-Spector(2011)Weiner and Grill-Spector]{facebody}
Kevin Weiner and Kalanit Grill-Spector.
\newblock Neural representations of faces and limbs neighbor in human
  high-level visual cortex: Evidence for a new organization principle.
\newblock \emph{Psychological research}, 77, 12 2011.
\newblock \doi{10.1007/s00426-011-0392-x}.

\bibitem[Weiner et~al.(2014)Weiner, Golarai, Caspers, Chuapoco, Mohlberg,
  Zilles, Amunts, and Grill-Spector]{mfs}
Kevin~S. Weiner, Golijeh Golarai, Julian Caspers, Miguel~R. Chuapoco, Hartmut
  Mohlberg, Karl Zilles, Katrin Amunts, and Kalanit Grill-Spector.
\newblock The mid-fusiform sulcus: A landmark identifying both
  cytoarchitectonic and functional divisions of human ventral temporal cortex.
\newblock \emph{NeuroImage}, 84:\penalty0 453--465, 2014.

\bibitem[Welling et~al.(2003)Welling, Osindero, and
  Hinton]{welling2003learning}
Max Welling, Simon Osindero, and Geoffrey~E Hinton.
\newblock
  \href{http://papers.nips.cc/paper/2177-learning-sparse-topographic-representations-with-products-of-student-t-distributions.pdf}{Learning
  sparse topographic representations with products of student-t distributions}.
\newblock In \emph{Advances in neural information processing systems}, pp.\
  1383--1390, 2003.

\bibitem[william Clemons(2018)]{body}
william Clemons.
\newblock Human body identification and verification dataset.
\newblock In \emph{Mendeley Data}, December 2018.

\bibitem[Yamins \& DiCarlo(2016)Yamins and DiCarlo]{yamins2016}
Daniel L~K Yamins and James~J DiCarlo.
\newblock Using goal-driven deep learning models to understand sensory cortex.
\newblock \emph{Nature Neuroscience}, 19\penalty0 (3):\penalty0 356--365, 2016.
\newblock \doi{10.1038/nn.4244}.
\newblock URL \url{https://doi.org/10.1038/nn.4244}.

\bibitem[Zhang et~al.(2021)Zhang, Zhou, Bao, and Liu]{VTCSOM}
Yiyuan Zhang, Ke~Zhou, Pinglei Bao, and Jia Liu.
\newblock Principles governing the topological organization of object
  selectivities in ventral temporal cortex.
\newblock \emph{bioRxiv}, 2021.
\newblock \doi{10.1101/2021.09.15.460220}.
\newblock URL
  \url{https://www.biorxiv.org/content/early/2021/09/17/2021.09.15.460220}.

\bibitem[Zhang et~al.(2017)Zhang, Song, and Qi]{zhifei2017cvpr}
Zhifei Zhang, Yang Song, and Hairong Qi.
\newblock Age progression/regression by conditional adversarial autoencoder.
\newblock \emph{CoRR}, abs/1702.08423, 2017.
\newblock URL \url{http://arxiv.org/abs/1702.08423}.

\bibitem[Zhou et~al.(2017)Zhou, Lapedriza, Khosla, Oliva, and
  Torralba]{zhou2017places}
Bolei Zhou, Agata Lapedriza, Aditya Khosla, Aude Oliva, and Antonio Torralba.
\newblock Places: A 10 million image database for scene recognition.
\newblock \emph{IEEE Transactions on Pattern Analysis and Machine
  Intelligence}, 2017.

\end{thebibliography}
\bibliographystyle{svrhm_2021}

\newpage
\appendix

% \section{Appendix}
\section{Additional Results}
\label{appendix:additional_results}

\subsection{Robustness to Initialization}
\label{appendix:init_robustness}
\begin{figure}[h!]
\centering
\includegraphics[width=1.0\linewidth]{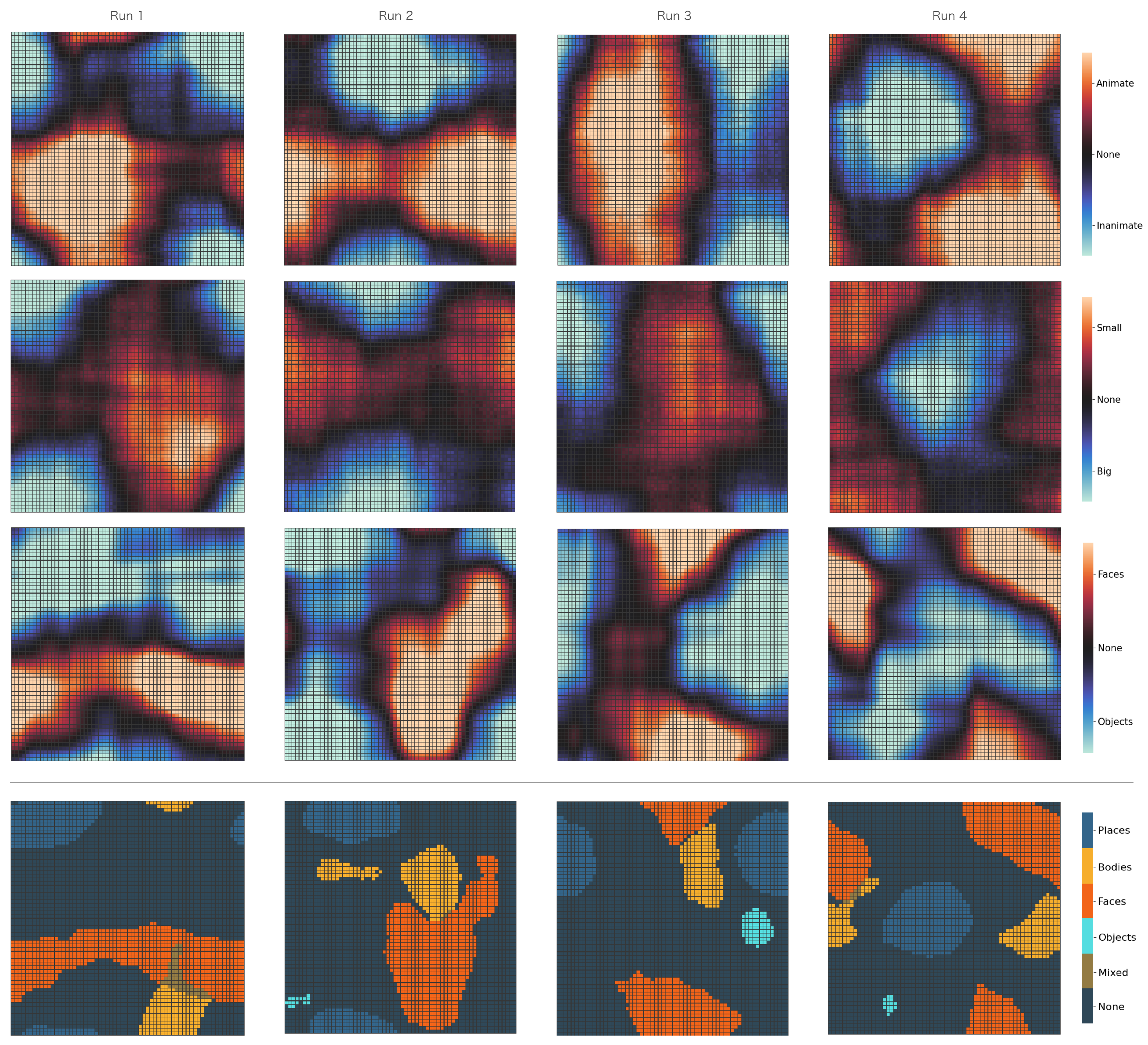}
\caption{Selectivity maps for the TVAE across four random initalizations. We observe that the emergent feature hierarchy and the relative placement of category-clusters is consistent in each case.} 
\label{fig:all_robustness}
\end{figure}

To verify the robustness of our results to randomness between trials, in Figure \ref{fig:all_robustness} below we compare the selectivity maps shown in the main text across four independant random initalizations of the weights. We first note that the emergent feature hierarchy depicted in Figure \ref{fig:hierarchy} appears roughly consistent across each trial. Specifically, selectivity to places, `big', and `inanimate' objects appears highly overlapping in each setting. We further note that the relative placement and size of the category-selective clusters (shown in the bottom row) is again roughly consistent across runs, with face and body clusters always adjacent and frequently overlapping. We see that in some runs, a small cluster selective to a generic `object' category can be observed. The relative weakness of this cluster is likely due to the lack of uniquely identifying features shared across all images in the object dataset. 

\subsection{Robustness to Face Test-Dataset Choice}
\label{appendix:data_robustness}
To investigate the robustness of face selectivity across different face test-datasets, and ensure the observed clusters are not a dataset dependant phenomenon, selectivity maps computed using four different face test-datasets are shown for both the TVAE and TDANN in Figure \ref{fig:data_robustness} below. Explicitly, the four datasets included: a ~25,000 subset of VGGface2 \ncite{cao2018vggface2}, 10,137 images from UTKface \ncite{zhifei2017cvpr}, 24,684 images from CelebA  \ncite{liu2015faceattributes}, and the Labled Faces in the Wild \cite{LFWTech} dataset upon which the models were trained. The resulting selectivity maps can be seen to be highly consistent despite the variability between low-level dataset statistics, indicating the observed selectivity is more likely related to the high level category information as desired.

\begin{figure}[h!]
\centering
\includegraphics[width=1.0\linewidth]{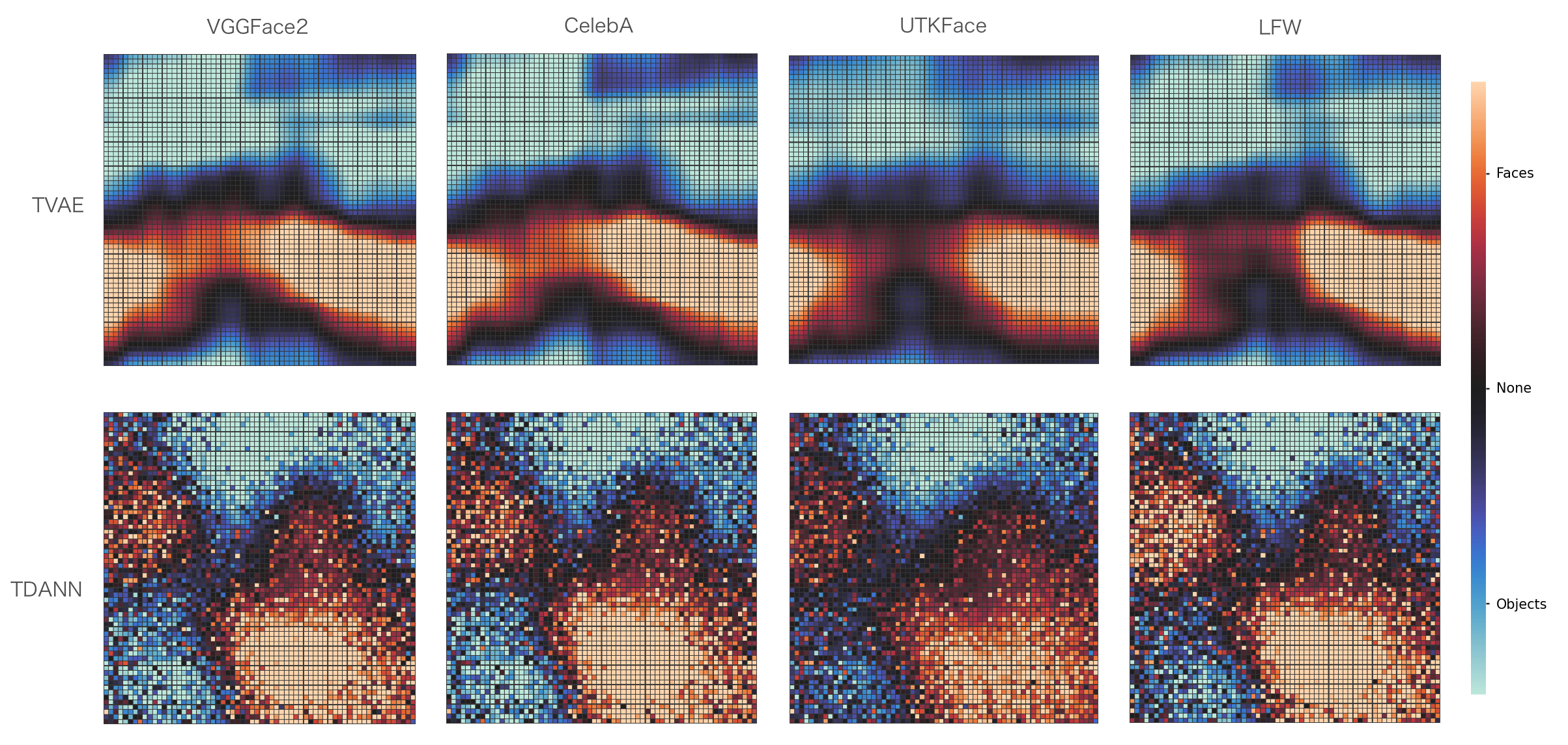}
\caption{Face vs. Object selectivity maps for four different face datasets. We see that for both the TVAE and TDANN the relative locations and sizes of the face and object selective clusters are stable despite the differences in the underlying test-datasets used.} 
\label{fig:data_robustness}
\end{figure}

\subsection{Distance-dependant Pairwise Correlation} 
\begin{figure}[h!]
\centering
\includegraphics[width=1.0\linewidth]{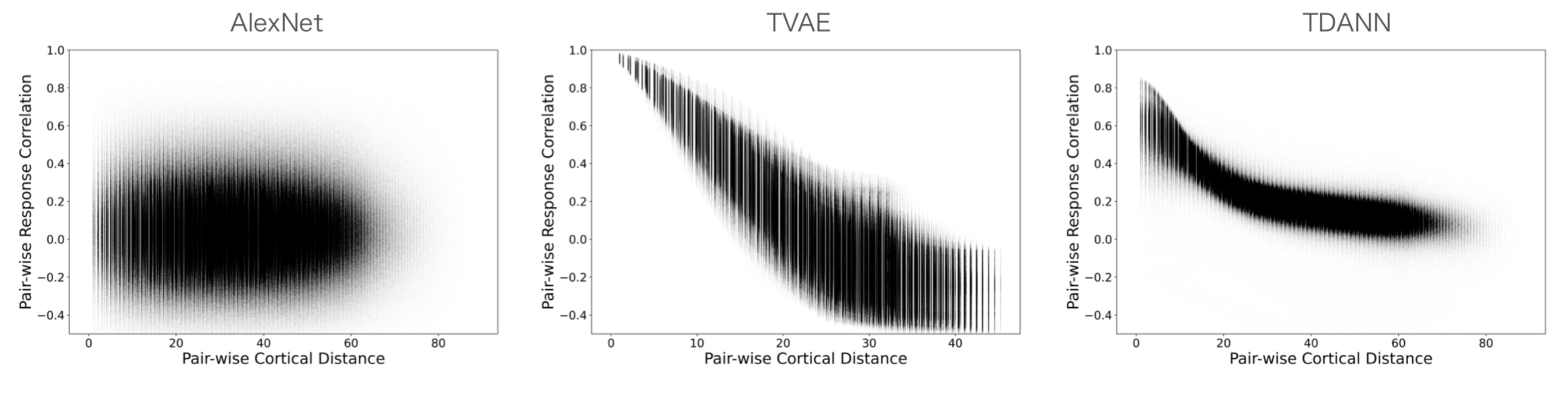}
% \vspace{-5mm}
\caption{Pairwise correlation between neurons as a function of distance in the cortical sheet.} 
\label{fig:pairwise_corr}
\end{figure}

To further quantify the topographic organization of the TVAE and how it compares with that of the TDANN, we measure the pairwise correlation (Pearson's R) of all topographic neurons as a function of distance in Figure \ref{fig:pairwise_corr}. We see that the TDANN (right) curve matches the original results \cite{TDANN}, roughly achieving the minimal spatial correlation loss, and mimicking the observed correlation curve from recordings in monkeys, as designed (see \cite{TDANN} for further discussion). Interestingly, the TVAE (middle) yields a qualitatively similar curve, despite having no such goal in its initial design. Finally, the correlation of the baseline model (left) is independant of distance as expected. We note that due to the circular boundary conditions of the TVAE, the maximal distance between neurons is significantly less, and thus scale of the X-axis is different between these two plots. In future work a more detailed comparison would benefit from matching boundary conditions in both models. Finally, in Figure \ref{fig:ksize} we plot the correlation curves for TVAEs trained with different spatial window sizes. We see that this has a significant effect on the shape of the curve, potentially allowing for more precise tuning to match biological data.

\subsection{Impact of TVAE Spatial Window Size ($\mathbf{W}$)}
In Figure \ref{fig:ksize} below, we demonstrate the effect of different choices of topographic organization (defined by $\mathbf{W}$) on the resulting learned selectivity maps. Specifically, we keep the global topography the same (a 2-d grid with circular boundary conditions), but we change the spatial extent over which variance is shared between variables $\mathbf{t}$. From left to right, we defined the matrix $\mathbf{W}$ to be a convolution matrix with kernels of size $5\times5$, $15\times15$, $25\times25$, and $35\times35$, where the total grid size is $64\times64$. 
\begin{figure}[h!]
\centering
\includegraphics[width=1.0\linewidth]{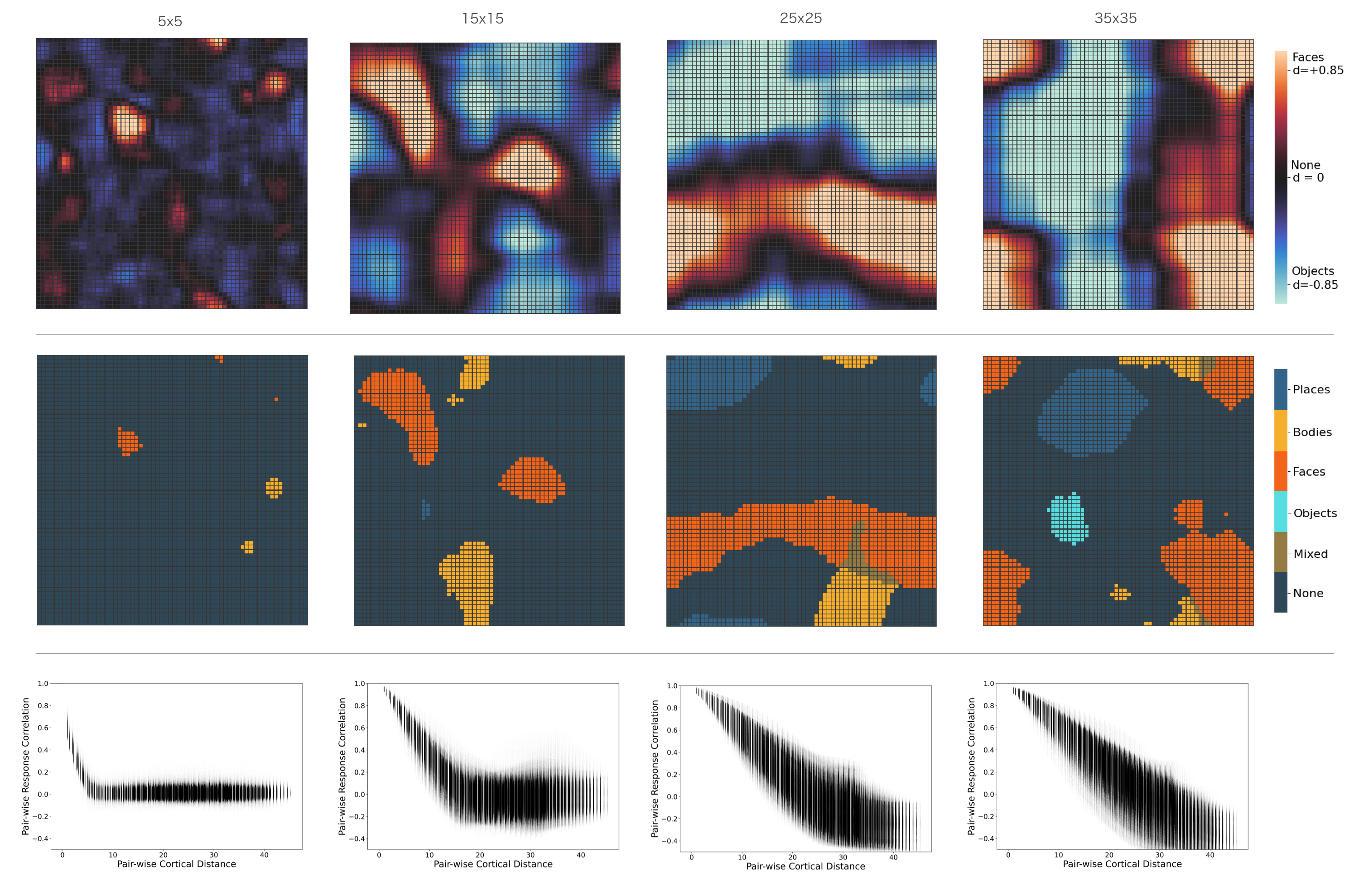}
\vspace{-5mm}
\caption{Selectivity maps and pairwise correlation curves for  different choices of spatial window size in the Topographic VAE.} 
\label{fig:ksize}
\end{figure}

\subsection{TDANN Nested Spatial Hierarchy}
In Figure \ref{fig:tdann_nested} below, we show the abstract selectivity maps for the TDANN, analogous to those in Figure \ref{fig:hierarchy} for the TVAE in the main paper. We see that the TDANN does appear to have a similar nested spatial hierarchy, however it is difficult to measure the differences visually. In future work, we hope to explore methods for quantifying the coherence of selectivity hierarchies, allowing greater comparison of models on this front. 
\begin{figure}[h!]
\centering
\includegraphics[width=1.0\linewidth]{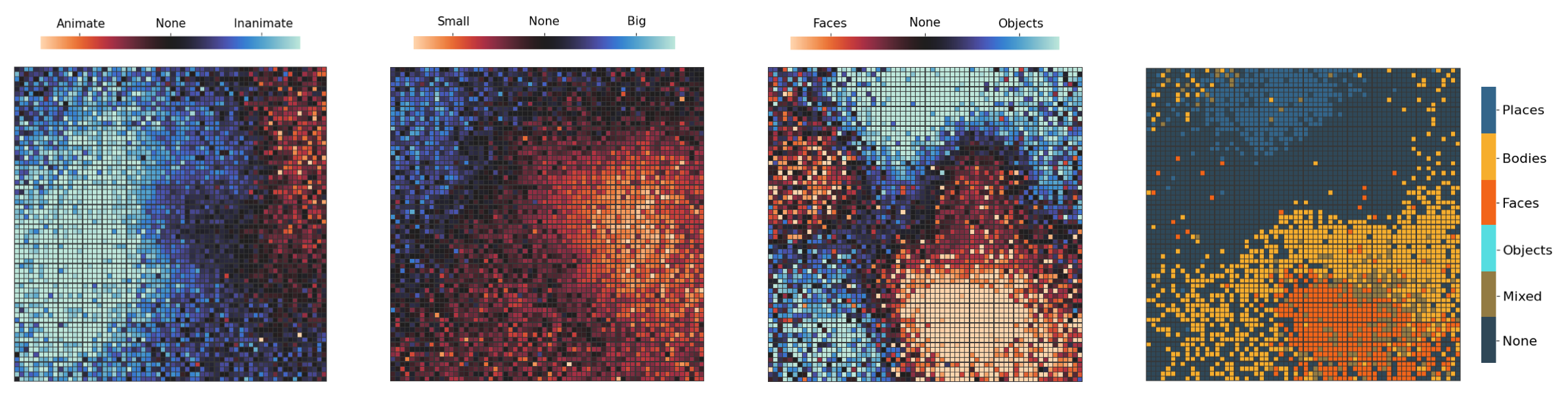}
\caption{Abstract category selectivity for the TDANN, analogous to the results presented in Figure \ref{fig:hierarchy} for the TVAE. From left to right: Animate vs. Inanimate, Small vs. Big, Faces vs. Objects, and  Multi-class selectivity with $d \geq 0.85$ (analagous to Figure \ref{fig:multiclass}).} 
\label{fig:tdann_nested}
\end{figure}

\subsection{VAE Baseline}
As an additional non-topographic baseline, we train a standard VAE in-place of the TVAE and measure the selectivity and single-image activation maps as in Figures \ref{fig:selectivity} and \ref{fig:single_img_maps}. Interestingly, we see that the standard VAE exhibits significantly fewer class-selective neurons, with the majority of neurons activating for each image. We find this correlates with the measured likelihood of the data under each model, suggesting that topographic organization (and similarly class-selectivity) acts as regularization on model performance, slightly reducing the overall likelihood. As measured in prior work \cite{leavitt2020selectivity}, high class-selectivity is similarly seen to be slightly detrimental to classification performance, agreeing with these results.
\begin{figure}[h!]
\centering
\includegraphics[width=1.0\linewidth]{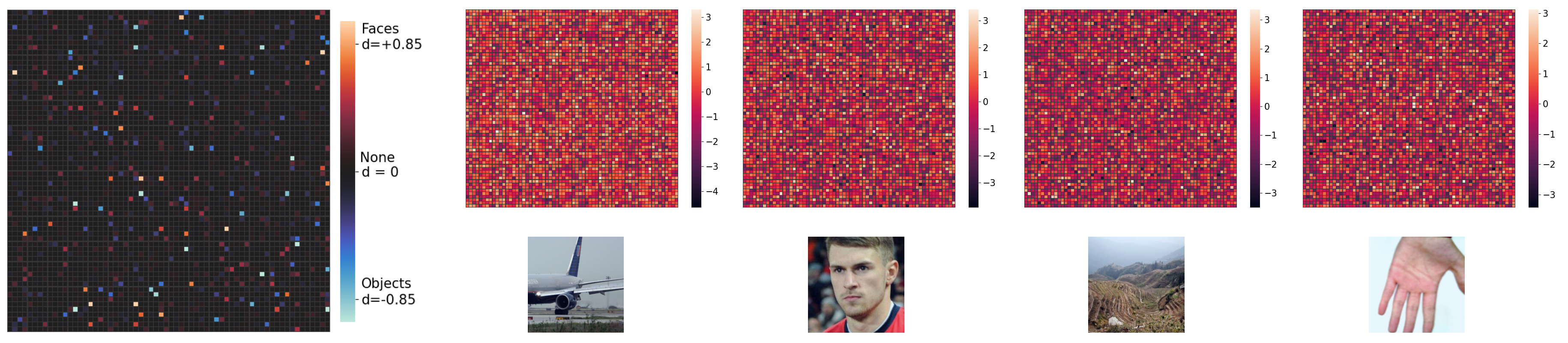}
\caption{Face vs. Object selecitivity (left) and single-image activation maps (right) for a non-topographic VAE baseline} 
\label{fig:vae_baseline}
\end{figure}

\section{Experimental details}
All code for running the experiments in this paper can be found at the following repository:\\ \url{https://github.com/akandykeller/CategorySelectiveTVAE}

\subsection{Training details}
\label{appendix:model_details}
\paragraph{Dataset Preprocessing} 
In order to eliminate variability between different datasets, all images were first reshaped to $256\times256$. A random percentage of the image area (between $8\%$ to $100\%$) and a random aspect ratio (between $\frac{3}{4}$ and $\frac{4}{3}$) were then chosen, and each image was then cropped according to these values. Finally, the crops were resized to the final shape of $224\times224$. All images were then normalized by the mean $[0.48300076, 0.45126104, 0.3998704]$ and standard deviation $[0.26990137, 0.26078254, 0.27288908]$.

\paragraph{TDANN Hyperparameters} The TDANN model was trained with stochastic gradient descent, a learning rate of $1 \times 10^{-3}$, standard momentum of $0.9$, and a batch size of $128$ for 10 epochs. Explicitly, the loss function was given by a sum of the classification cross entropy loss, the spatial correlation losses for both layers FC6 and FC7, and weight decay of $5 \times 10^{-4}$. A fixed weight of $10 \times \frac{1}{4096^2}$ was multiplied by the spatial correlation loss before backpropagating as this was found necessary to qualitatively match the results from \citet{TDANN}. Contrary to the original TDANN work, we did not randomly initialize the locations of the neurons, and instead spaced them evenly on a grid of the same size. We found the spatial correlation loss to still function equally well in this setting, and detail our implementation in Section \ref{appendix:spatial_loss} below.

\paragraph{TVAE Hyperparameters} The TVAE was trained with stochastic gradient descent, a learning rate of $1 \times 10^{-5}$, standard momentum of $0.9$, and a batch size of $128$ for 30 epochs. The global topology was set to a single 2D torus (i.e. a 2D grid with circular boundary conditions), and the local topology was set to sum of local regions of size $25\times25$, i.e. the kernel used to convolve over $\mathbf{u}$ was of size $25\times25$ and contained all $1$'s. The $\mu$ parameter was initialized to $40$, and trained simultanously with the remainder of the model parameters. 

\subsection{Spatial Correlation Loss of TDANN}
\label{appendix:spatial_loss}
The exact form of the spatial correlation loss used for training the TDANN in this paper is given as:
\begin{equation}
    \mathrm{SpatialCorrelationLoss}(\mathbf{z}) = \sum_i^n \sum^n_{j \neq i} \left|C_{ij}(\mathbf{z}) - \frac{1}{D_{ij} + 1} \right|
\end{equation}
where $\mathbf{z}$ an $n$-dimensional vector of activations, $C$ is the normalized cross correlation matrix (e.g. a matrix of Pearson correlation coefficients), and D is a matrix containing the `cortical distances' in millimeters between all pairs of neurons $i$ and $j$. In this work, we defined all neurons to be equally spaced in a 2-D grid of 10mm $\times$ 10mm. This resulted in a horizontal and vertical spacing between neurons of $0.15625$mm and a diagonal spacing of $0.22097087$mm. Unlike the TVAE, the TDANN grid was not defined to have circular boundary conditions in order to match the original model.

% \subsection{Body \& Place Clusters in TDANN}

% \subsection{Spatial Nested Hierarchy in TDANN}

% \input{sections/relatedwork}

\end{document}